\documentclass[11pt]{article}

\usepackage{multirow,amssymb,amsbsy,amsmath}

\usepackage{graphicx}
\usepackage{verbatim}
\usepackage{booktabs}
\usepackage{dcolumn}
\usepackage{float}

\usepackage{authblk}
\usepackage{color}

\usepackage{bm}

\numberwithin{equation}{section}

\author[1]{Johannes Nokkala\thanks{jsinok@utu.fi}}
\author[1,2]{Sabrina Maniscalco}
\author[1]{Jyrki Piilo}

\affil[1]{Turku Centre for Quantum Physics, Department of Physics and Astronomy,
University of Turku, FI-20014, Turun Yliopisto, Finland}
\affil[2]{Centre for Quantum Engineering, Department of Applied Physics, School of Science,
Aalto University, P.O. Box 11000, FIN-00076 Aalto, Finland}

\begin{document}

\title{Non-Markovianity over ensemble averages in quantum complex networks}

\maketitle

\begin{abstract}
We consider bosonic quantum complex networks as structured finite environments for a quantum harmonic oscillator and  investigate the interplay between the network structure and its spectral density, excitation transport properties and non-Markovianity. After a review of the formalism used, we demonstrate how even small changes to the network structure can have a large impact on the transport of excitations. We then consider the non-Markovianity over ensemble averages of several different types of random networks of identical oscillators and uniform coupling strength. Our results show that increasing the number of interactions in the network tends to suppress the average non-Markovianity. This suggests that tree networks are the random networks optimizing this quantity.
\end{abstract}

\section{Introduction}
\setcounter{equation}{0}

Understanding the dynamics of open quantum systems is important in several fields of physics and chemistry including problematics dealing, e.g., with quantum to classical transition and decoherence with its harmful effects for quantum information processing and communication. In general, formulating or deriving a suitable equation of motion for the density matrix plöt for the open system is often a daunting task. Perhaps the most celebrated and most used theoretical result in this context is the Gorini-Kossakowski-Sudarshan-Lindblad (GKSL) master equation \cite{GKS76,L76}

\begin{equation}
\dfrac{d \rho_s(t)}{dt}=-i[H_s,\rho_s(t)]+\sum_k \gamma_k \left( C_k\rho_s(t) C_k^\dagger-\dfrac{1}{2} \left \{ C_k^\dagger C_k,\rho_s(t) \right \} \right),
\end{equation}

\noindent with the associated completely positive and trace preserving dynamical map with semigroup property. Above, $H_s$ is the open system Hamiltonian, $\gamma_k$ are positive constant rates, and $C_k$ are the jump operators with $k$ indexing the different decoherence channels. Indeed, this master equation and the corresponding publications had recently 40th anniversary celebrations in the Symposium on Mathematical Physics in Toru{\'n} in June 2016. 

GKSL master equation $(1.1)$ describes Markovian memoryless open system dynamics and during the last 10-15 years there has been an increasing amount of research activities in understanding memory effects and quantifying non-Markovianity for open systems beyond the semigroup property \cite{Rivas14,Vega17,Breuer16}. A pair of complementary approaches here include a description based on quantifying the information flow between the open system and its environment \cite{Breuer09} or the characterization of dynamical maps in terms of their divisibility properties \cite{Rivas14,Ch14} while a large number of other ways to characterize non-Markovianity also exist, see e.g. \cite{Wolf08,Lu10,Luo12,Lorenzo13,Bylicka14}. Most of the research so far has focused on non-Markovianity using discrete variable open systems as examples while in the current work we are interested in the memory effects in a continuous variable (CV) open system with controlled environmental structure.

Indeed, here we consider structured finite environments modeled by bosonic quantum complex networks. While this and other kinds of quantum complex networks have recieved increasing attention in recent years in the context of perfect state transfer \cite{Christandl04,Yung05}, quantum random walks \cite{Kempe10,Faccin13}, efficient entanglement distribution \cite{Cirac97,Acin07,Cuquet09} and the unification of classical and quantum network theory \cite{Bianconi15,Biamonte17}, here the focus is on the interplay between the network structure and the reduced dynamics of an open quantum system attached to it. To this end, we investigate the impact of the structure on the network spectral density, excitation transport properties and non-Markovianity of the reduced dynamics.

The paper is organized as follows. Section 2 concerns the network itself. Here we present the microscopic model and briefly discuss the connection between the network Hamiltonian and certain matrix representations of abstract graphs in classical graph theory. The dynamics of the network is given in terms of a symplectic matrix acting on the vector of operators at initial time. In Section 3, we describe how complex quantum networks can be treated in the framework of the theory of open quantum systems as tunable structured environments. We demonstrate how small changes in the network structure can have a large impact on its excitation transport properties. In Section 4, we consider the non-Markovianity of the reduced dynamics using a recently introduced witness based on non-monotonicity of the evolution of Gaussian interferometric power. Finally, conclusions are drawn in Section 5. 

\section{Bosonic quantum complex networks}
\setcounter{equation}{0}

\subsection{The Hamiltonian}

We set $\hbar=1$ and work with position and momentum operators defined as $q=(a^\dagger+a)/\sqrt{2\omega}$ and $p=(a^\dagger-a)i\sqrt{\omega/2}$, satisfying the commutation relation $[q,p]=i$. We consider networks of $N$ unit mass quantum harmonic oscillators coupled by springlike couplings. The general form of a Hamiltonian for such networks is

\begin{equation}
H_E=\dfrac{\mathbf{p}^T\mathbf{p}}{2}+\mathbf{q}^T\mathbf{Aq},
\end{equation}

\noindent where we have introduced the vectors of position and momentum operators $\mathbf{q}^T=\{q_1,...,q_N\}$ and $\mathbf{p}^T=\{p_1,...,p_N\}$, and where $\mathbf{A}$ is the matrix containing the coupling terms and frequencies. It has elements $\mathbf{A}_{ij}=\delta_{ij}\tilde{\omega}_i^2/2-(1-\delta_{ij})g_{ij}/2$, where $g_{ij}$ is the strength of the springlike coupling $g_{ij}(q_i-q_j)^2/2$ between the position operators of oscillators $i$ and $j$, and $\tilde{\omega}_i^2=\omega_i^2+\sum_j g_{ij}$ is the effective frequency of oscillator $i$ resulting from absorbing the quadratic parts of the coupling terms into the free Hamiltonians of the oscillators.

The matrix $\mathbf{A}$, which completely determines the network Hamiltonian, can be related to some of the typical matrix representations of weighted graphs, i.e. abstract networks of nodes connected by weighted edges. By weighted, we mean that a magnitude is assigned to each connection. This can be used to establish a link between the properties of the network and results from graph theory. A paradigmatic example is the adjacency matrix $\mathbf{V}$ having elements $\mathbf{V}_{ij}=w_{ij}$, where $w_{ij}$ is the weigth of the connection between nodes $i$ and $j$; a weigth of 0 corresponds to the nodes being disconnected.  Another matrix that arises very naturally is the Laplace matrix $\mathbf{L}$, related to the adjacency matrix as $\mathbf{L}=\mathbf{D}-\mathbf{V}$, where $\mathbf{D}$ is diagonal with elements $\mathbf{D}_{ii}=\sum_j w_{ij}$. In terms of them, matrix $\mathbf{A}$ can be written as $\mathbf{A}=\mathbf{\Delta}_{\tilde{\omega}}^2/2-\mathbf{V}/2$ or as $\mathbf{A}=\mathbf{\Delta}_{\omega}^2/2+\mathbf{L}/2$, where $\mathbf{\Delta}_{\tilde{\omega}}$ and $\mathbf{\Delta}_{\omega}$ are diagonal matrices of the effective and bare frequencies of the network oscillators, respectively, and weights are given by the coupling strengths. The graph aspect of this and other kinds of quantum networks have been very recently used to, e.g., develop a local probe for the connectivity and coupling strength of a quantum complex network by using results of spectral graph theory \cite{Nokkala17a}, and constructing Bell-type inequalities for quantum communication networks by mapping the task to a matching problem of an equivalent unweighted bipartite graph \cite{Luo17}.

The Hamiltonian $(2.1)$ is a special case of the quadratic Hamiltonian $H=\mathbf{x}^T\mathbf{Mx}$, where the vector $\mathbf{x}$ contains both the position and momentum operators and $\mathbf{M}$ is a $2N \times 2N$ matrix such that $H$ is Hermitian. It can be shown \cite{Ticochinsky79} that quadratic Hamiltonians can be diagonalized to arrive at an equivalent eigenmode picture of uncoupled oscillators provided that $\mathbf{M}$ is positive definite. Since $H$ is Hermitian, this is equivalent with the positivity of the eigenvalues of $\mathbf{M}$. In the case at hand, $H_E$ may be diagonalized with an orthogonal matrix $\mathbf{K}$ such that $\mathbf{K}^T\mathbf{AK}=\mathbf{\Delta}$, where the diagonal matrix $\mathbf{\Delta}$ holds the eigenvalues of $\mathbf{A}$. By defining new operators

\begin{equation}
\begin{cases} 
\mathbf{Q}=\mathbf{K}^T\mathbf{q} \\ 
\mathbf{P}=\mathbf{K}^T\mathbf{p}, \\
\end{cases}
\end{equation}

\noindent the diagonal form of $H_E$ reads

\begin{equation}
H_E=\dfrac{\mathbf{P}^T\mathbf{P}}{2}+\mathbf{Q}^T\mathbf{\Delta}\mathbf{Q},
\end{equation}

\noindent which is the Hamiltonian of $N$ decoupled oscillators with frequencies $\Omega_i=\sqrt{2\mathbf{\Delta}_{ii}}$.


\subsection{The dynamics of the network}

A bosonic quantum complex network is also an interesting system to study in its own right. Below, we review the mathematical tools useful for the task, adopting the definitions for a commutator and anti-commutator between two operator-valued vectors used in \cite{Menicucci11}. While we will be later concerned with networks initially in the thermal state, we will also briefly discuss the case of an initial Gaussian state without displacement. For a more detailed review of Gaussian formalism in phase space, see \cite{Ferraro05}. What is presented here is straightforward to apply to the case where interactions with external oscillators is considered, and we will do so in Section $3$.  

Let $\mathbf{x}$ be a vector containing the position and momentum operators of the network oscillators, and define the commutator between two operator valued vectors as $[\mathbf{x}_1,\mathbf{x}_2^T]=\mathbf{x}_1\mathbf{x}_2^T-(\mathbf{x}_2\mathbf{x}_1^T)^T$. Now canonical commutation relations give rise to a symplectic form $\mathbf{J}$, determined by $[\mathbf{x},\mathbf{x}^T]=i\mathbf{J}$. Let $\mathbf{x}'=\mathbf{S}\mathbf{x}$, where $\mathbf{S}$ is a $2N \times 2N$ matrix of real numbers. In order for $\mathbf{S}$ to be a canonical transformation of $\mathbf{x}$, the commutation relations must be preserved. This requirement gives $i\mathbf{J}=[\mathbf{x}',\mathbf{x}'^T]=[\mathbf{S}\mathbf{x},(\mathbf{S}\mathbf{x})^T]=\mathbf{S}[\mathbf{x},\mathbf{x}^T]\mathbf{S}^T=i\mathbf{S}\mathbf{J}\mathbf{S}^T$, implying that $\mathbf{SJS}^T=\mathbf{J}$. Such a matrix is called symplectic with respect to symplectic form $\mathbf{J}$. Symplectic matrices form the symplectic group $Sp(2N,\mathbb{R})$ with respect to matrix multiplication, which can be used to define a symplectic representation of the Gaussian unitary group, meaning that (up to an overall phase factor) the two groups are bijective. 

We fix $\mathbf{x}^T=\{\mathbf{q}^T,\mathbf{p}^T\}=\{q_1,...,q_N,p_1,...,p_N\}$ throughout the rest of the present work. Then the symplectic form becomes $\mathbf{J}=\left(\begin{smallmatrix}
0 & \mathbf{I}_N \\
-\mathbf{I}_N & 0
\end{smallmatrix}\right)$, where $\mathbf{I}_N$ is the $N\times N$ identity matrix. By defining the vector of eigenmode operators to be $\mathbf{X}^T=\{\mathbf{Q}^T,\mathbf{P}^T\}=\{Q_1,...,Q_N,P_1,...,P_N\}$, we can express the transformation that diagonalizes the network Hamiltonian as $\mathbf{X}=\left(\begin{smallmatrix}
\mathbf{K}^T & 0 \\
0 & \mathbf{K}^T
\end{smallmatrix}\right)\mathbf{x}$; a direct calculation shows that the matrix diagonalizing the Hamiltonian is both symplectic and orthogonal.

In the eigenmode picture, the equations of motion are those of noninteracting oscillators. By defining the auxiliary diagonal matrices with elements $\mathbf{D}_{\cos ii}^\Omega=\cos(\Omega_i t)$, $\mathbf{D}_{\sin ii}^\Omega=\sin(\Omega_i t)$ and $\mathbf{\Delta}_{\Omega ii}=\Omega_i$, we can express them as

\begin{equation}
\begin{pmatrix}
\mathbf{Q}(t) \\
\mathbf{P}(t)
\end{pmatrix}  = 
\begin{pmatrix}
	\mathbf{D}_{\cos}^\Omega & \mathbf{\Delta}_{\Omega}^{-1}\mathbf{D}_{\sin}^\Omega \\
	-\mathbf{\Delta}_{\Omega}\mathbf{D}_{\sin}^\Omega & \mathbf{D}_{\cos}^\Omega
\end{pmatrix}
\begin{pmatrix}
\mathbf{Q}(0) \\
\mathbf{P}(0)
\end{pmatrix},
\end{equation}

\noindent where the block matrix acting on the vectors is again symplectic. To recover the dynamics of the network oscillators, we may use Eq. $(2.2)$ to express $\mathbf{x}(t)$ in terms of either $\mathbf{X}(0)$ as

\begin{equation}
\begin{pmatrix}
\mathbf{q}(t) \\
\mathbf{p}(t)
\end{pmatrix}  = 
\begin{pmatrix}
	\mathbf{K}\mathbf{D}_{\cos}^\Omega & \mathbf{K}\mathbf{\Delta}_{\Omega}^{-1}\mathbf{D}_{\sin}^\Omega \\
	-\mathbf{K}\mathbf{\Delta}_{\Omega}\mathbf{D}_{\sin}^\Omega & \mathbf{K}\mathbf{D}_{\cos}^\Omega
\end{pmatrix}
\begin{pmatrix}
\mathbf{Q}(0) \\
\mathbf{P}(0)
\end{pmatrix},
\end{equation}

\noindent or in terms of $\mathbf{x}(0)$ as

\begin{equation}
\begin{pmatrix}
\mathbf{q}(t) \\
\mathbf{p}(t)
\end{pmatrix}  = 
\begin{pmatrix}
	\mathbf{K}\mathbf{D}_{\cos}^\Omega\mathbf{K}^T & \mathbf{K}\mathbf{\Delta}_{\Omega}^{-1}\mathbf{D}_{\sin}^\Omega\mathbf{K}^T \\
	-\mathbf{K}\mathbf{\Delta}_{\Omega}\mathbf{D}_{\sin}^\Omega\mathbf{K}^T & \mathbf{K}\mathbf{D}_{\cos}^\Omega\mathbf{K}^T
\end{pmatrix}
\begin{pmatrix}
\mathbf{q}(0) \\
\mathbf{p}(0)
\end{pmatrix}.
\end{equation}

\noindent Notice that the group properties of symplectic matrices quarantees that in both cases the block matrix remains symplectic.

If we now restrict our attention to Gaussian states with zero mean, we may define the covariance matrix of the initial state as

\begin{equation}
\mathrm{cov}(\mathbf{x}(0))=\tfrac{1}{2}\langle[\mathbf{x}(0),\mathbf{x}^T(0) ]_+\rangle,
\end{equation}

\noindent where the anti-commutator is defined as $[\mathbf{x}_1,\mathbf{x}_2 ]_+=\mathbf{x}_1\mathbf{x}_2^T+(\mathbf{x}_2\mathbf{x}_1^T)^T$. If $\mathbf{x}(t)=\mathbf{S}\mathbf{x}(0)$, then the covariance matrix at time $t$ becomes

\begin{equation}
\begin{aligned}
\mathrm{cov}(\mathbf{x}(t)) & =\mathrm{cov}(\mathbf{S}\mathbf{x}(0))=\tfrac{1}{2}\langle[(\mathbf{S}\mathbf{x}(0),(\mathbf{S}\mathbf{x}(0))^T ]_+\rangle \\
& =\tfrac{1}{2}\mathbf{S}\langle[\mathbf{x}(0),\mathbf{x}(0))^T ]_+\rangle\mathbf{S}^T=\mathbf{S}\mathrm{cov}(\mathbf{x}(0))\mathbf{S}^T.
\end{aligned}
\end{equation}

In the present case of symplectic matrices appearing in Eqs. $(2.5)$ and $(2.6)$, the choice depends on the basis where the initial covariance matrix is defined. A particular subtlety concerns an initial thermal state for the network, where either choice might seem natural. Here, assuming the usual thermal expectation values for non-interacting oscillators in the real oscillator basis, i.e. a diagonal $\mathrm{cov}(\mathbf{x}(0))$, corresponds to the case where the interactions are suddenly switched on at $t=0+$. As here the state is not the stationary state with respect to the Hamiltonian $(2.1)$, one will see the excitations of each network oscillator evolve with time. On the other hand, if one assumes the covariance matrix to be diagonal in the eigenmode basis instead, the excitations will be frozen. In this work we are using the latter approach as it is quite natural to assume an initial stationary state for the environment of an open quantum system. 

While here the correlation structure in the state of the network is not studied, it is of great interest in the emerging field of continuous-varibale quantum information processing and in particular in the study of so-called cluster states \cite{Menicucci06,Zhang06}, which are multi-mode correlated states used as a resource in measurement-based quantum computing. In this context, it is typically the state, rather than the Hamiltonian, that is represented with a graph. It has been shown that specific quadratic Hamiltonians have cluster states as their ground state, which can then be adiabatically prepared by cooling a set of non-interacting modes to zero temperature and then switching on the interactions \cite{Aolita11}.

Finally, we mention the complementary viewpoint of open quantum networks, where the network is considered as the open system interacting with an environment of infinite size. The dynamics can then be described with a master equation for the network density matrix. Collective phenomena, such as synchronization, can occur in a network relaxing towards a steady state \cite{Manzano13}.

\subsection{Experimental aspect}

To implement an oscillator network, the basic requirements to meet are a static topology, harmonic potential
and quantum regime for the oscillators. To match the form of the Hamiltonian $(2.1)$, the couplings between the oscillator position operators should be springlike, and any other interactions between them should either be eliminated or minimized. 

More challenging requirements include the scalability to many nodes and the ability to implement also long-range couplings in order to have a nontrivial topology. The biggest difficulties are related to the implementation of generic networks: essentially a platform reconfigurable to a desired static topology would be needed, i.e. independent control and tunability over all couplings would be necessary.

A possible way to implement a simple oscillator network is to use vibrational modes of trapped ions. In this
way, it is possible to implement simple oscillator chains that interact in a harmonic way via Coulomb force in single
or segmented traps \cite{ions1,ions2}. The main limitations are related to scalability and independent control of couplings. In
particular, if the couplings are mediated by Coulomb force, then they cannot be controlled in an independent way,
which limits the networks that can be realized in this way. Proposals for scalable arrays of trapped ions have
been made \cite{ionsproposal}.

One can also consider cold atoms trapped in optical lattices. They offer a scalable platform to simulate different many-body systems, in particular the Bose-Hubbard Hamiltonian, which describes interacting bosons in a lattice. While the Hamiltonian is different, it still shares some similarities with that of an oscillator network. The parameters of the Hamiltonian can be tuned, but it cannot be used to implement an arbitrary topology.

An array of coupled micro- or nanomechanical resonators acting as phonon traps is a natural candidate for an experimental realization. The setup has good scalability, as experimental implementations of arrays of up to 400 resonators have been
reported \cite{resonatorarray}. In the case of mechanically coupled devices, independent control of the coupling strengths might not be
possible, however a proposal of a fully reconfigurable resonator array based on optical couplings has been made \cite{reconfigurablearray}.
Other challenges include the suppression of intrinsic nonlinearities of the devices, as well as cooling them to reduce
thermal noise. First steps in this direction have been taken, as coherent phonon manipulation has been reported in
a system of two resonators with a tunable mechanical coupling \cite{coherentphonons}.

Perhaps the most promising alternative is the very recently proposed optical implementation of the dynamics given by the  Hamiltonian $(2.1)$, based on a simultaneous downconversion of the components of an optical frequency comb from a femtosecond laser followed by pulse shaping and mode-selective measurements \cite{Nokkala17b}. By mapping the Hamiltonian to quadrature operators of the optical field modes and determining the so called Bloch-Messiah decomposition of either the symplectic matrix $(2.5)$ or $(2.6)$, one will find the pulse shape and measurement basis necessary to implement it. In particular, since the network structure is mapped into the parameters of the platform, changing the network does not require a change in the optical setup. The result is a deterministic and highly reconfigurable implementation of quantum complex networks with in principle arbitrary structure. In practice, producing the required pump shape to a sufficiently good accuracy will require further theoretical and experimental work before the proposal can be tested.

\section{Quantum networks as structured environments}
\setcounter{equation}{0}

\subsection{Attaching external oscillators}

We consider as the open quantum system a single additional quantum harmonic oscillator interacting with one of the network oscillators. While this is sufficient to our present purposes, what follows is straightforward to extend to the case of multiple external oscillators or interactions with multiple network nodes. Moreover, we will fix the states of the open system and the network to be a Gaussian state and a thermal state of temperature $T$, respectively, assume factorizing initial conditions and work in such units that the Boltzmann constant $k_B=1$.

The open system Hamiltonian is $H_S=(p_S^2+\omega_S^2q_S^2)/2$, and the form of the interaction Hamiltonian reads $H_I=-kq_Sq_i$, or equivalently, $H_I=-kq_S\sum_j^N \mathbf{K}_{ij}Q_j$ in the basis of eigenmodes, where $k$ is the coupling strength between the open system and the network. The total Hamiltonian is now $H=H_S+H_E+H_I$. By including the operators of the open system as the final elements of the vectors of operators, we may express it analogously to Hamiltonian $(2.1)$ as

\begin{equation}
	H=\dfrac{\{\mathbf{P},p_S\}^T\{\mathbf{P},p_S\}}{2}+\{\mathbf{Q},q_S\}^T\mathbf{B}\{\mathbf{Q},q_S\},
\end{equation}

\noindent where the matrix $\mathbf{B}$ has diagonal elements $\mathbf{B}_{ii}=\Omega_i^2/2$ for $i<N+1$ and $\mathbf{B}_{N+1,N+1}=\omega_S^2/2$, while $\mathbf{B}_{N+1,i}=\mathbf{B}_{i,N+1}=-k\mathbf{K}_{li}/2$ for $i<N+1$; here the index $l$ is the index of the network oscillator directly interacting with the open system. We may diagonalize the matrix $\mathbf{B}$ as $\mathbf{O}^T\mathbf{B}\mathbf{O}=\mathbf{F}$ where $\mathbf{O}$ is orthogonal and $\mathbf{F}$ diagonal with elements $F_{ii}=f_i^2/2$, where $f_i$ will be the frequencies of the modes in the fully diagonal picture. If we define the new operators as

\begin{equation}
\begin{cases} 
\bm{\mathcal{Q}}=\mathbf{O}^T \{\mathbf{Q},q_S\}\\ 
\bm{\mathcal{P}}=\mathbf{O}^T \{\mathbf{P},p_S\}, \\
\end{cases}
\end{equation}

\noindent the total Hamiltonian reads

\begin{equation}
H_E=\dfrac{\bm{\mathcal{P}}^T\bm{\mathcal{P}}}{2}+\bm{\mathcal{Q}}^T\mathbf{F}\bm{\mathcal{Q}}.
\end{equation}

We are now in position to write down the symplectic matrix giving the dynamics of the total Hamiltonian. By following the steps leading from Hamiltonian $(2.3)$ to Eq. $(2.6)$, we arrive at 

\begin{equation}
\begin{pmatrix}
\mathbf{Q}(t) \\
q(t)\\
\mathbf{P}(t) \\
p(t)\\
\end{pmatrix}  = 
\begin{pmatrix}
	\mathbf{O}\mathbf{D}_{\cos}\mathbf{O}^T & \mathbf{O}\mathbf{\Delta}_{f}^{-1}\mathbf{D}_{\sin}\mathbf{O}^T \\
	-\mathbf{O}\mathbf{\Delta}_{f}\mathbf{D}_{\sin}\mathbf{O}^T & \mathbf{O}\mathbf{D}_{\cos}\mathbf{O}^T
\end{pmatrix}
\begin{pmatrix}
\mathbf{Q}(0) \\
q(0)\\
\mathbf{P}(0) \\
p(0)\\
\end{pmatrix},
\end{equation}

\noindent where we have introduced the diagonal matrices  $\mathbf{D}_{\cos ii}=\cos(f_i t)$, $\mathbf{D}_{\sin ii}=\sin(f_i t)$ and $\mathbf{\Delta}_{f ii}=f_i$.

As we will consider an initial thermal state for the network, throughout the rest of the present work we will consider as the initial basis the one on the R.H.S. of the equation above, where the initial covariance matrix of the network is diagonal with elements $\langle Q_i(0)^2 \rangle=(n_i+1/2)/\Omega_i$ and $\langle P_i(0)^2 \rangle=(n_i+1/2)\Omega_i$, where $n_i=(\exp(\Omega_i/T)-1)^{-1}$. 

If we are interested in the dynamics of the operators in the network basis, we may use Eq. $(2.2)$ and define the symplectic and orthogonal $N+1 \times N+1$ matrix $\tilde{\mathbf{K}}$ with elements $\tilde{\mathbf{K}}_{N+1,i}=\mathbf{K}_{i,N+1}=0$ for $i<N+1$, $\tilde{\mathbf{K}}_{N+1,N+1}=1$, and $\tilde{\mathbf{K}}_{ij}=\mathbf{K}_{ij}$ otherwise. Now

\begin{equation}
\begin{pmatrix}
\mathbf{q}(t) \\
q(t)\\
\mathbf{p}(t) \\
p(t)\\
\end{pmatrix}  = 
\begin{pmatrix}
	\tilde{\mathbf{K}}\mathbf{O}\mathbf{D}_{\cos}\mathbf{O}^T & \tilde{\mathbf{K}}\mathbf{O}\mathbf{\Delta}_{f}^{-1}\mathbf{D}_{\sin}\mathbf{O}^T \\
	-\tilde{\mathbf{K}}\mathbf{O}\mathbf{\Delta}_{f}\mathbf{D}_{\sin}\mathbf{O}^T & \tilde{\mathbf{K}}\mathbf{O}\mathbf{D}_{\cos}\mathbf{O}^T
\end{pmatrix}
\begin{pmatrix}
\mathbf{Q}(0) \\
q(0)\\
\mathbf{P}(0) \\
p(0)\\
\end{pmatrix}.
\end{equation}

\noindent The dynamics can be readily determined from the initial covariance matrix of the total system, as outlined in Eq. $(2.8)$, where the result will be the covariance matrix at time $t$ in the basis of either the eigenmodes or the network oscillators, depending on which symplectic matrix is used. If the open system has displacement, also the evolution of its first moments needs to be considered to determine the evolution of its state.

While we are concerned with the dynamics of the open system as well as the network oscillators, we mention here the possibility to treat the network in the framework of Gaussian channels. For a general Gaussian state and for $\mathbf{x}(0)=\left(\begin{smallmatrix}q(0) \\p(0) \end{smallmatrix}\right)$, the elements of the covariance matrix of a single mode system are $\mathrm{cov}(\mathbf{x}(0))_{ij}=\langle \mathbf{x}(0)_i \mathbf{x}(0)_j+\mathbf{x}(0)_j \mathbf{x}(0)_i \rangle/2-\langle \mathbf{x}(0)_i \rangle\langle \mathbf{x}(0)_j\rangle$. For any Gaussian channel taking the covariance matrix to time $t$, the transformation can be written as

\begin{equation}
\mathrm{cov}(\mathbf{x}(t))=\mathbf{C}(t)\mathrm{cov}(\mathbf{x}(0))\mathbf{C}(t)^T+\mathbf{L}(t),
\end{equation}

\noindent where $\mathbf{C}(t)$ and $\mathbf{L}(t)$ are real matrices and $\mathbf{L}(t)$ is symmetric. In terms of the elements of the symplectic matrix $\mathbf{S}$ of Eq. $(3.4)$, we may find the elements using Eq. $(2.8)$ to be

\begin{equation}
\mathbf{C}(t)=\begin{pmatrix}\mathbf{S}_{N+1,N+1}&\mathbf{S}_{N+1,2N+2}\\\mathbf{S}_{2N+2,N+1}&\mathbf{S}_{2N+2,2N+2}\end{pmatrix},
\end{equation}

\noindent and

\begin{equation}
\mathbf{L}(t)=\sum_i \langle \mathbf{X}_i(0)^2 \rangle
\begin{pmatrix}
\mathbf{S}_{N+1,i}^2 & \mathbf{S}_{N+1,i}\mathbf{S}_{2N+2,i}\\
\mathbf{S}_{N+1,i}\mathbf{S}_{2N+2,i} & \mathbf{S}_{2N+2,i}^2
\end{pmatrix},
\end{equation}

\noindent where the sum is taken to $2N+1$ excluding $N+1$, such that $\mathbf{L}(t)$ is independent of the initial expectation values of the open system. The matrices $\mathbf{C}(t)$ and $\mathbf{L}(t)$ now completely characterize the channel, allowing, e.g. to make comparisons with channels defined by a master equation or to construct intermediate channels taking the system from time $t>0$ to $s>t$ and checking if the resulting channel is completely positive or not, as is done in a recently introduced measure of non-Markovianity for Gaussian channels \cite{Torre15}. The difficulty in implementing this measure in the present case is neither in the construction of the intermediate map nor checking its complete positivity, but rather in the fact that it considers the limit $s \rightarrow t$, and it is not clear how to take such a limit in the case of numerical, rather than analytical, matrices $\mathbf{C}(t)$ and $\mathbf{L}(t)$.

\subsection{The spectral density}

One of the central concepts in the theory of open quantum systems is the spectral density of environmental couplings $J(\omega)$, which encodes the relevant information in the environment and interaction Hamiltonians into a single function of frequency. The reduced dynamics of the open system can then be determined once the initial state of the total system as well as the system Hamiltonian are fixed \cite{Weiss}. In particular, a heat bath is completely characterized by its spectral density and temperature. The definition of the spectral density, in terms of the environment eigenfrequencies $\Omega_i$ and coupling strengths to eigenmodes $g_i$, reads

\begin{equation}
J(\omega)=\dfrac{\pi}{2}\sum_i\dfrac{g_i^2}{\Omega_i}\delta(\omega-\Omega_i),
\end{equation}

\noindent where $\delta$ is the Dirac's delta function. The definition is rarely used in practice, since in the case of an infinite heat bath with a continuum of frequencies the spectral density becomes a continuous function, and phenomenological spectral densities are defined instead.

In the case of finite environments it is convenient to use the relation between $J(\omega)$ and the damping kernel $\gamma(t)$, the latter appearing in the generalized quantum Langevin equations giving the dynamics for the open system operators \cite{Weiss}. It is defined as

\begin{equation}
\gamma(t)=\sum_i\dfrac{g_i^2}{\Omega_i^2}\cos(\Omega_i t),
\end{equation}

\noindent and the relation is given by

\begin{equation}
J(\omega)=\omega\int_0^\infty\gamma(t)\cos(\omega t) dt,
\end{equation}

If the environment is finite, both Eq. $(3.9)$ and Eq. $(3.11)$ will result in delta spikes. However, by replacing the upper limit of integration by a finite time $t_{max}$, the intermediate form of the spectral density can be considered instead. If a quantum network defined by a Hamiltonian of the form $(2.1)$ is sufficiently symmetric, the reduced dynamics will have a regime where the system interacts with a continuum of frequencies as if the environment was infinite. This is evident from the damping kernel having a very small value during this transient, until finite size effects cause a revival of oscillations. The duration of this continuous regime of reduced dynamics depends on the structure and size of the finite environment.

\begin{figure*}[h!]
                \includegraphics[trim=0cm 0cm 0cm 0cm,clip=true,width=0.95\textwidth]{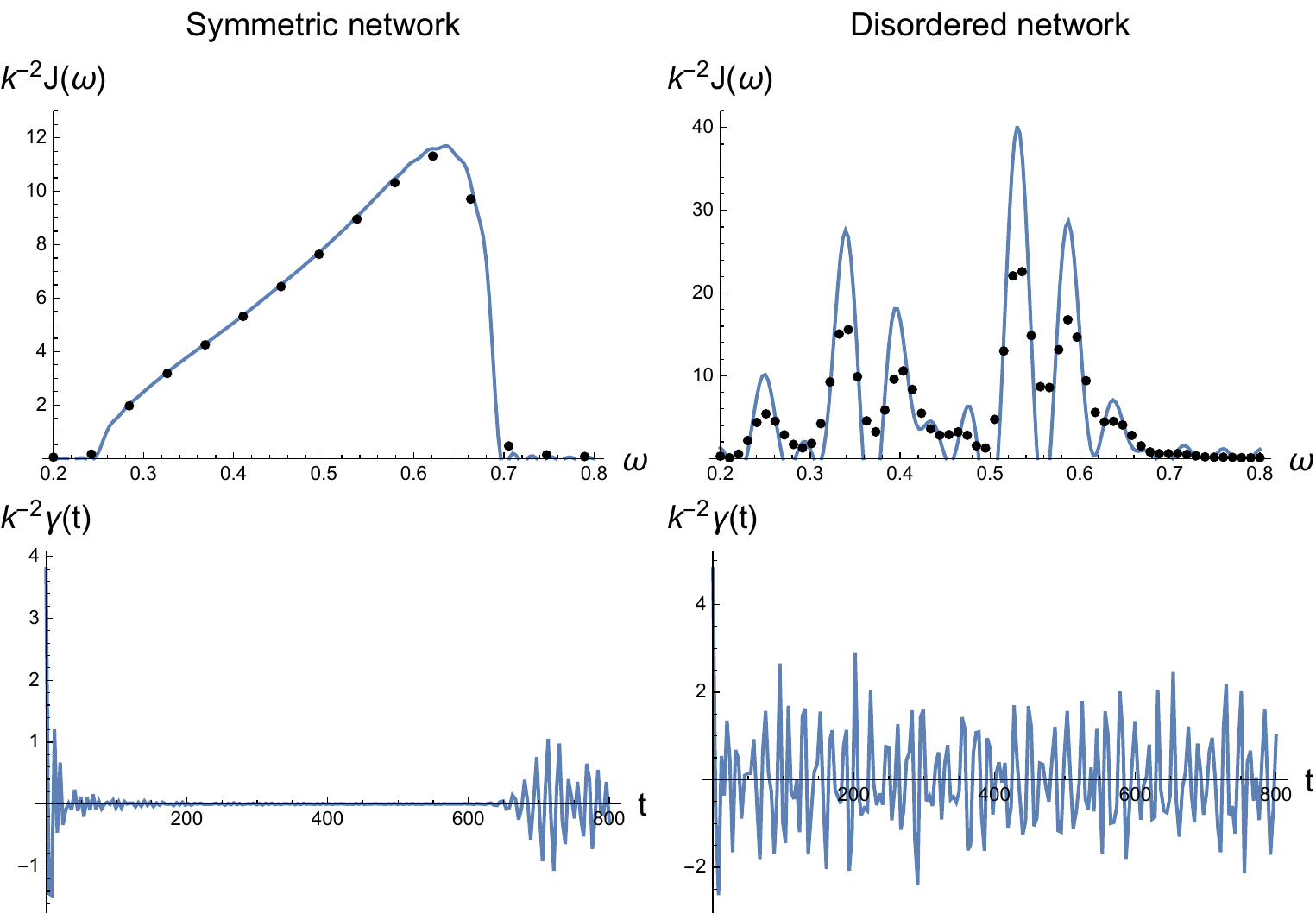}
            \caption{\label{figJY}
(Color online) A comparison of the spectral densities and damping kernels for a symmetric and a disordered network. The black dots are probed values of $J(\omega)$ extraced from the reduced dynamics of an open quantum system interacting with the networks. The symmetric network is a chain of $N=100$ oscillators with nearest and next nearest neighbor couplings with magnitudes of $g_1=0.1$ and $g_2=0.02$, respectively, while the disordered network is a random network of $N=30$ oscillators with a constant coupling strength $g=0.05$. For both, the bare frequency of the oscillators was $\omega_0=0.25$, the system-network interaction strength was $k=0.01$ and the states of the system and the network where a thermal state of $T=1$ and vacuum, respectively. The system is coupled to the first oscillator in the chain and to a random oscillator of the disordered network.
}
      \end{figure*}

In the present case of quantum complex networks, the coupling strengths to eigenmodes $g_i$ are determined by the interaction Hamiltonian $H_I$ and the matrix $\mathbf{K}$ diagonalizing the Hamiltonian $(2.1)$ as $g_i=-k \mathbf{K}_{li}$, where $l$ is the index of the network oscillator directly interacting with the system and $k$ the interaction strength in the network basis. In Fig. \ref{figJY}, we show two examples of damping kernels and spectral densities for quantum networks. The symmetric network is a chain with nearest and next nearest couplings. Additionally, the chain is made homogeneous by setting the effective frequencies of the ends of the chain equal with the rest. The spectral density is continuous for the used value of $t_{max}$. If the interaction time is sufficiently short, it would not be possible to tell from the reduced dynamics of an open quantum system coupled to the network alone that the environment is in fact finite. In contrast, the disorder in the other network results in a highly structured spectral density that does not have a continuous regime.

In general, it may be asked whether $J(\omega)$ of a quantum complex network can be deduced from the reduced dynamics of the system. It can be shown \cite{Nokkala16} that, provided the coupling to the network $k$ is weak and the network is in a thermal state, the system excitation number is well approximated by the expression $\langle n(t) \rangle=\exp(-\Gamma t)\langle n(0) \rangle+n(\omega_S)(1-\exp(-\Gamma t)$, where $\Gamma=J(\omega_S)/\omega_S$ and $n(\omega_S)=(\exp(\omega_S/T)-1)^{-1}$, or the thermal average boson number at system frequency $\omega_S$. The value of the spectral density at system frequency is then approximated by

\begin{equation}
J(\omega_S)=\dfrac{\omega_S}{t}\ln\left( \dfrac{\Delta n(0)}{\Delta n(t)} \right),
\end{equation}

\noindent where $\Delta n(t)=n(\omega_S)-\langle n(t) \rangle$. If $T$ is known, the local value of the spectral density can be determined by performing measurements on the system only. This is demonstrated in Fig. \ref{figJY}, where the dots are probed values of the spectral density with each circle corresponding to one value of the system frequency. By keeping the interaction time fixed to the used value of $t_{max}$, it can be seen that even for networks with disorder, the probed values follow the shape of $J(\omega)$.  

It is also worth mentioning that the machinery introduced so far can be used to approximate an infinite heat bath, determined by its spectral density, with a finite one. Together with its temperature, the finite bath is completely characterized by the coupling strengths $g_i$ and frequencies $\Omega_i$. While there is considrebale freedom in choosing $\Omega_i$, they should cover the non-vanishing parts of $J(\omega)$ and there should be enough of them to push the finite size effects to interaction times longer than what is being considered. Next, the couplings are determined from the spectral density as follows. From Eq. $(3.9)$, it can be seen that $\int_0^\infty\frac{2}{\pi}J(\omega)\omega d\omega=\sum_i g_i^2$. Approximating the integral on the left hand side with, e.g., a Riemann sum, and identifying the terms on both sides then gives $g_i^2=\frac{2}{\pi}J(\Omega_i)\Omega_i\Delta\Omega_i$, where $\Delta\Omega_i=|\Omega_i-\Omega_{i+1}|$ is the sampling interval. The range of interaction times where the approximation is valid can be checked by comparing the damping kernels calculated for the finite bath from Eq. $(3.10)$ and for the infinite bath from the inversion of Eq. $(3.11)$, namely, $\gamma(t)=\frac{2}{\pi}\int_0^\infty\frac{J(\omega)}{\omega}\cos(\omega t)d\omega$. The two will be similar up to the point where finite size effects manifest. This can be of advantage when considering early or intermediate dynamics in the case of a strong coupling, since the dynamics given by Eqs. $(3.4)$ or $(3.5)$ is exact.

\subsection{Engineering  aspect and excitation transport}

Reservoir engineering aims to modify the properties of the environment of an open quantum system, typically to protect non-classicality of the system or to increase the efficiency of some task. In the present case, the environment is a quantum network determined by the matrix $\mathbf{A}$. To assess its properties as an environment, it is convenient to consider the effect of the structure on the spectral density $J(\omega)$, which can be returned to the effect of the structure on the eigenfrequencies $\Omega_i$ and coupling strengths to eigenmodes $g_i$. By changing the structure by, e.g., adding or removing links, one can try to effectively decouple the system from the network by finding a configuration where $J(\omega_S)$ has a small value, or alternatively to look for structures with increases transport efficiency.  

In fact, assuming that the system can be freely coupled to any single node in the network, a single network can produce as many spectral densities as it has nodes. This is because a coupling to a single node corresponds to a set of coupling strengths $g_i$, which are in turn directly proportional to a row of the matrix $\mathbf{K}$ diagonalizing the network. On the other hand, the set of eigenfrequencies $\Omega_i$ are completely determined by the eigenvalues of the matrix $\mathbf{A}$ and as such are independent of where in the network the system is coupled. 

\begin{figure*}[t]
                \includegraphics[trim=0cm 0cm 0cm 0cm,clip=true,width=0.95\textwidth]{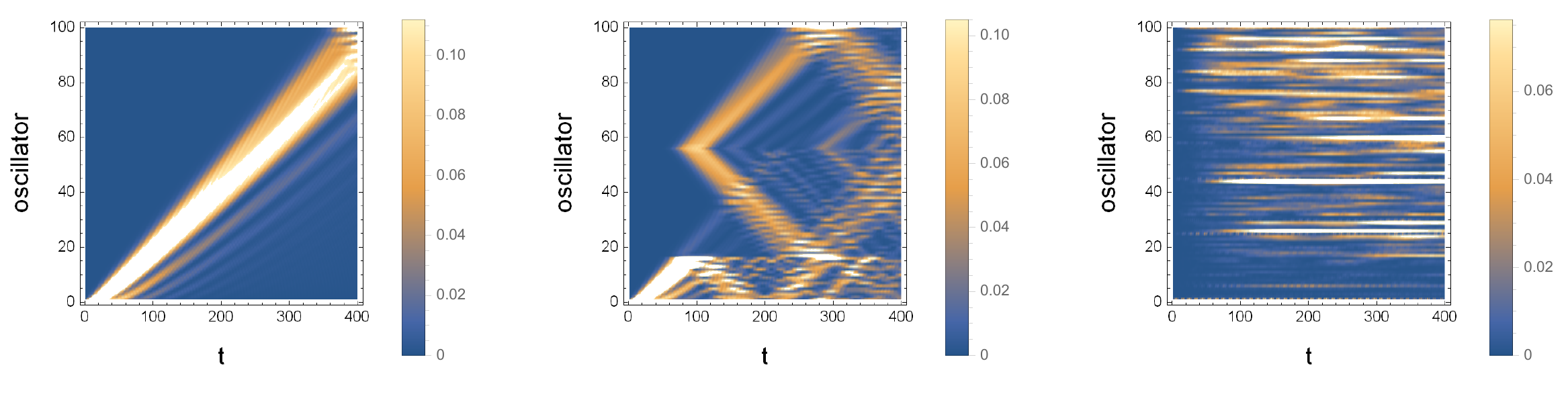}
            \caption{\label{figtransport}
(Color online) Examples of excitation transport in quantum networks. In all examples, the interaction time is shown on the horizontal axis while the vertical axis corresponds to the index of the network oscillator. The color bar shows the difference between initial excitations and excitations at time $t$. On the left, the network is the symmetric network of Fig. \ref{figJY}. Excitations propagate freely along the chain. In the middle, a single randomly chosen link in the symmetric network has been rewired, changing the transport properties. On the right, evolution of excitations in the network oscillators of a random network of $N=100$ oscillators with bare frequency $\omega_0=0.25$ and coupling strengths $g=0.05$ is shown. Excitations become locked to a subset of network oscillators. 
}
      \end{figure*}

Even small changes to the network structure can have a large impact on both the network spectral density and excitation transport properties. Generally speaking, when the reduced dynamics has a continuous regime, the flow of energy is steady provided that the system is resonant with the network. Furthermore, excitations can freely be exchanged between different nodes in the network. On the other hand, when the degree of disorder in the network is high, the excitations typically become locked to a subset of the network nodes and cannot spread effectively. We present examples of this in Fig. \ref{figtransport}, where the same symmetric network is considered as in  Fig. \ref{figJY}. Rewiring randomly only a single coupling changes the path taken by the majority of excitations. Also shown is the excitation dynamics in a random network.

While transport is inefficient in most random networks, a search can be carried out for exceptions, and indeed it can be shown that when sampling the distribution of random networks, some rare cases have vastly superior transport properties robust against ambient dephasing \cite{Scholak11}. One may also ask whether there is any connection between the excitation transport properties and non-Markovianity. While in the spin-boson model non-Markovianity and the back-flow of excitations can behave similarly with respect to the environment parameters \cite{Guarnieri16a}, there does not seem to be such a connection in the case of a continuous variable system \cite{Guarnieri16b}. Furthermore, even in the spin-boson model, information and excitation backflows can occur without the other \cite{Schmidt16}.

\section{Non-Markovianity in complex quantum networks}
\setcounter{equation}{0}

\subsection{Generalities}

The dynamics of an open quantum system can significantly deviate from the memoryless Markovian case when the interaction between the open system and the environment is strong, or if the environment is structured. Previous investigations \cite{Vasile14} of harmonic chains with nearest neighbor couplings, having a Hamiltonian of the form $(2.1)$, show that the strongest memory effects occur when the system frequency is located near the edges of the spectral density. A $J(\omega)$ with a single band will then have two regimes of system frequency where memory effects are strong while one with band-gaps will have more. In this work, the Breuer-Laine-Piilo \cite{BLP} and Rivas-Huelga-Plenio \cite{RHP} measures were used.   

To the best of our knowledge, however, there have been no studies of non-Markovianity attempting to connect it to the structure of a complex network. While it is the case that any spectral density of an oscillator network with non-regular structure can be replicated with an oscillator chain with nearest neighbor couplings \cite{linearmappingA,linearmappingB,linearmappingC}, it is nevertheless of interest to ask whether the amount of non-Markovianity could be tied to the statistical properties of complex networks by comparing the average non-Markovianity over many realizations, and whether adding more structure typically increases the non-Markovianity or not.

\begin{figure*}[t]
                \includegraphics[trim=0cm 0cm 0cm 0cm,clip=true,width=0.95\textwidth]{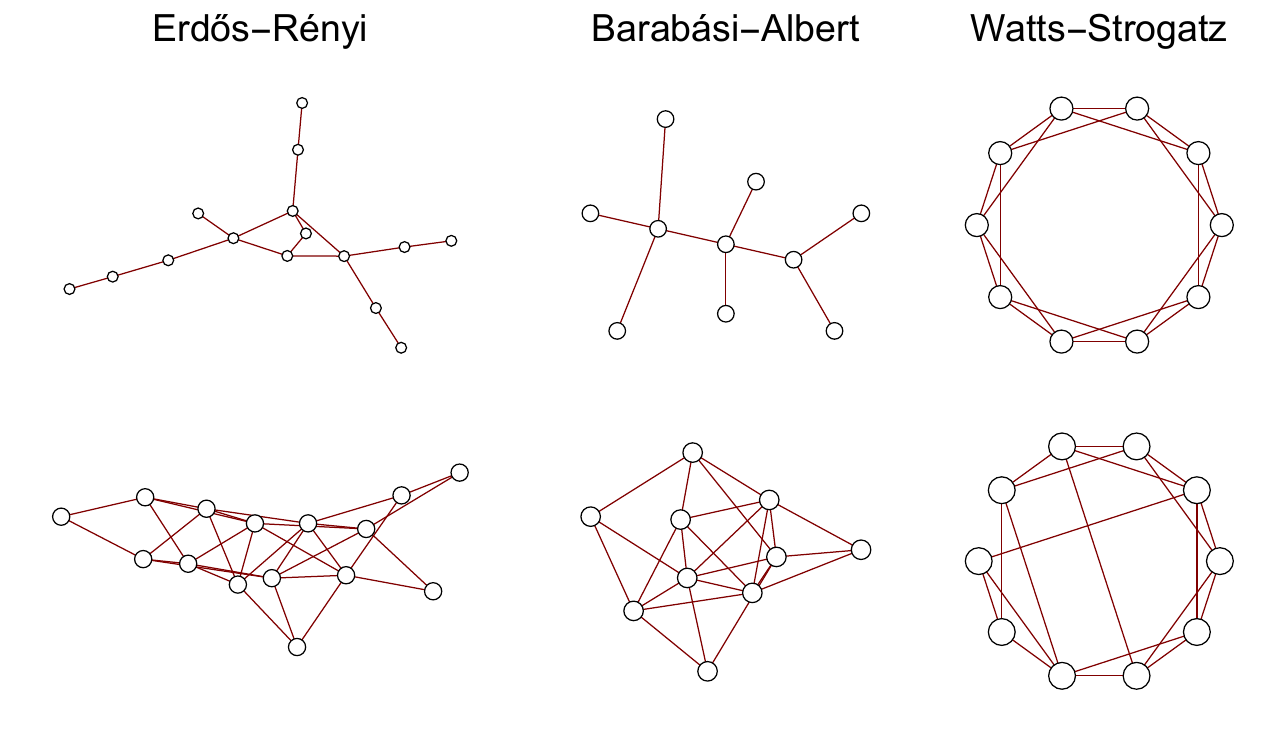}
            \caption{\label{figgraphs}
(Color online) Schematics for the used networks. Each column corresponds to a network type while the rows correspond to different parameter values. When the connection probability for the Erd\H{o}s-R\'{e}nyi network is increased the number of links grows, but links are chosen randomly. This is at variance with the Barab\'{a}si-Albert network where nodes with a higher number of links are preferred when introducing new links, resulting in highly connected nodes when the connectivity parameter grows. Watts-Strogatz networks are constructed from a cycle graph by rewiring each link with a given probability. As this rewiring probability grows, the average distance between the nodes decreases, but the total number of links remains constant.     
}
      \end{figure*}

To this end, we considered three types of random networks presented in Figure \ref{figgraphs}. For all three cases, we fixed the size of the network to be $N=30$ and assumed that the network is connected, i.e. any node can be reached from any other by following the links. The Erd\H{o}s-R\'{e}nyi network $G(N,p)$ \cite{ER} is constructed from the completely connected network of $N$ nodes by independently selecting each link to be part of the final network with a probability $p$. The Barab\'{a}si-Albert network $G(N,l)$ \cite{BA} is constructed from a connected network of $3$ nodes and repeatedly adding a new node with $l$ links, connecting it randomly to existing nodes but favoring nodes which already have a high number of links, until the size $N$ is reached. Setting $l=1$ is an important special case, as the resulting network is a tree, i.e. it has the smallest possible number of links that a connected network of size $N$ can have. Finally, a Watts-Strogatz network $G(N,p,n)$ \cite{WS} is constructed starting from a circular network where all nodes are connected to $n$-th nearest neighbors, and then rewiring each link with the probability $p$. In this work, we fixed $n=2$.

\subsection{Non-Markovianity quantified by the non-monotonicity of Gaussian interferometric power}

The key concept used in several witnesses and measures of non-Markovianity is to track the dynamics of a quantity that can be shown to behave differently under Markovian and non-Markovian evolutions. In this work, we consider a recently introduced measure and a witness based on the non-monotonicity of Gaussian interferometric power under non-divisible dynamical maps \cite{Souza15}.

Gaussian inteferometric power $\mathcal{Q}$ quantifies the worst-case precision achievable in black-box phase estimation using a bipartite Gaussian probe composed of modes $A$ and $B$. It is also a measure of discord-type correlations between the two modes, as it vanishes for product states. For quantifying non-Markovianity, it is enough to consider the case where mode $A$ is subjected to a local Gaussian channel while mode $B$ remains unchanged. Then the expression for the Gaussian interferometric power $\mathcal{Q}$ has a closed form in terms of the symplectic invariants of the two-mode covariance matrix $\sigma_{AB}$ \cite{Adesso14}.    

For Markovian channels, $\mathcal{Q}$ is a monotonically non-increasing function of time, implying that $\dfrac{d}{dt}\mathcal{Q}(\sigma_{AB})\leq 0$. Any period of time where this does not hold is then a sign of non-Markovianity. Once the initial covariance matrix $\sigma_{AB}$ has been fixed, the degree of non-Markovianity of the reduced dynamics can then be quantified as

\begin{equation}
\mathcal{N}_{GIP}=\frac{1}{2}\int_0^\infty (|\mathcal{D}(t)|+\mathcal{D}(t))dt,
\end{equation}

\noindent where $\mathcal{D}(t)=\dfrac{d}{dt}\mathcal{Q}(\sigma_{AB})$. While the related measure is defined with a maximization over all initial states for the bi-partite system, Eq. $(4.1)$ provides a lower bound for this measure. Since there is strong numerical evidence that squeezed thermal states are particularly suited for witnessing non-Markovianity of this type \cite{Souza15}, we fix the initial state of the two-mode system to be a squeezed thermal state with two-mode squeezing parameter $r=\frac{1}{2}\cosh^{-1}(5/2)$ and initial thermal excitations $n_A=n_B=1/2$ for both modes. The network initial state is taken to be the vacuum.  

Besides disorder in the network structure, additional sources of non-Markovianity include finite size effects that become stronger as the interaction time is increased, and memory effects at the boundaries of the spectral density. To better assess the non-Markovianity arising from the structure, we will restrict the interaction time to an intermediate value of $t=50$ and fix the frequency of the system to be the $15th$ eigenfrequency of the networks to ensure that it is resonant.

\begin{figure*}[h!]
                \includegraphics[trim=0cm 0cm 0cm 0cm,clip=true,width=0.95\textwidth]{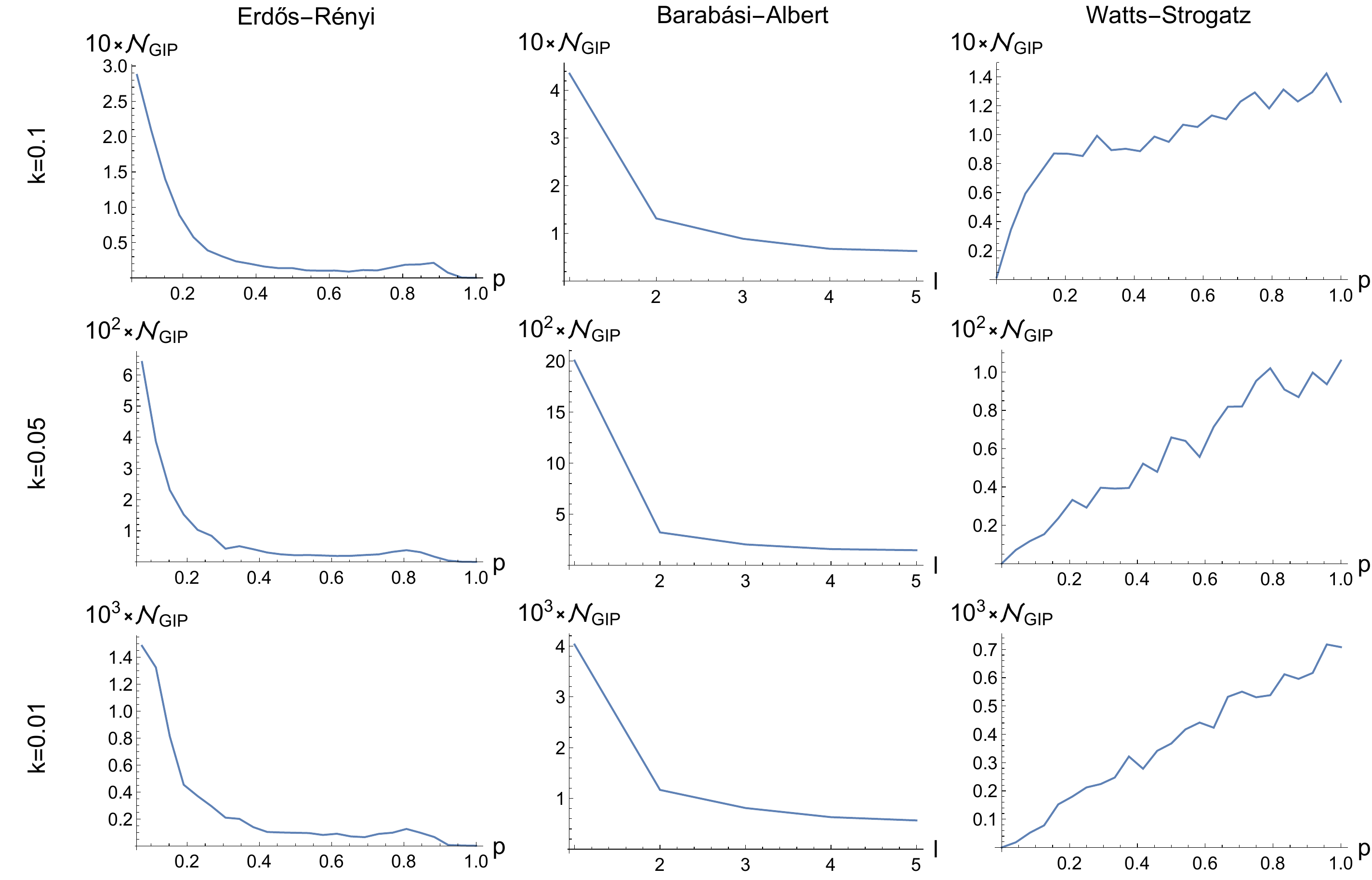}
            \caption{\label{figGIP}
(Color online) A comparison of the non-Markovianity for three different types of quantum complex networks. The columns correspond to the type and the rows to interaction strength between the network and the system. The size of each network is fixed to $N=30$ while a parameter controlling the structure of the network is varied. The parameters are connection probability $p$, connectivity parameter $l$ and rewiring probability $p$ for Erd\H{o}s-R\'{e}nyi, Barab\'{a}si-Albert and Watts-Strogatz networks, respectively. Refer to main text for details. Results are averaged over 1000 realizations for each parameter value.
}
      \end{figure*}

The results are shown in Fig. \ref{figGIP}. For all considered cases, changing the interaction strength affects the magnitude but not the behaviour of non-Markovianity against the network parameter. For Erd\H{o}s-R\'{e}nyi and  Barab\'{a}si-Albert networks, the number of couplings between network oscillators grows with the parameter, reducing the amount of non-Markovianity. On the other hand, the number of couplings in the network is constant for the Watts-Strogatz network. The results suggest that when the system is resonant with the network, non-Markovianity is highest for networks with a small amount of random couplings. For all considered coupling strengths, the highest non-Markovianity is achieved when the network is a tree. If the network is highly symmetric, as is the case with Watts-Strogatz networks with a low rewiring probability, the amount of non-Markovianity in the resonant case is very small. Non-Markovianity is increased by introducing disorder into the network through rewiring of the couplings.  

Besides the results we present here, we also checked that increasing the network temperature decreases the non-Markovianity. Furthermore, for comparison we determined the non-Markovianity in the simple case of a homogeneous chain with nearest-neighbor couplings only and found that even at the edges of the spectral density, where memory effects are strongest, $\mathcal{N}_{GIP}$ has a similar value than Erd\H{o}s-R\'{e}nyi and  Barab\'{a}si-Albert networks have in the resonant case. 

\section{Conlusions and outlook}
\setcounter{equation}{0}

In this work, we have studied bosonic quantum complex networks in the framework of open quantum systems. After briefly investigating the effect of the network stucture on the spectral density and transport of excitations, we focused on the non-Markovianity in the reduced dynamics of an open quantum system interacting with the network.

We considered non-Markovianity over ensemble averages of different types of random networks of identical oscillators and constant coupling strength between the network oscillators. Previous work shows that strong memory effects can occur in symmetric networks at the edges of the spectral density and near band gaps. Here we have shown that increasing the disorder of the network can lead to a high degree of non-Markovianity also when the system is resonant with the network, however increasing the number of interactions between network oscillators appears to suppress it, suggesting that trees optimize the ensemble averaged non-Markovianity.

While here we considered only the lower bound of a single non-Markovianity measure, it would be interesting to extend the investigations to other measures such as the measure introduced by Torre, Roga and Illuminati \cite{Torre15}. We expect that a systematic study could perhaps link some of the graph invariants, such as the mean distance between nodes, to non-Markovianity and other non-classical properties of the quantum networks, such as the ability to generate or transport entanglement. Such a link could pave way to structural control of non-classical properties of quantum complex networks. Indeed, in the case of quantum walks on classical complex networks, it can be shown that the quantumness of the walk is a function of both the initial state and specific graph invariants. Furthermore, for a deeper understanding of quantum networks the introduction of purely quantum graph invariants without a classical counterpart would be needed.     

\section*{Acknowledgments}
The authors acknowledge financial support from the Horizon 2020 EU collaborative projects QuProCS (Grant Agreenement No. 641277). J. N. acknowledges the Wihuri foundation for financing his graduate studies.



\end{document}